\documentstyle[prb,aps,epsfig]{revtex}

\begin{document}
\draft

 \title{Vibronic contributions to resonant NLO responses:\\
 two-photon absorption in push-pull chromophores}

 \author{Anna Painelli\thanks{Corresponding author: Dip. Chimica GIAF, 
Universit\`a di Parma, viale delle Scienze 17,
I--43100 Parma, Italy.
Phone: +39--0521--905461; fax: +39--0521--905556; 
e-mail: anna.painelli@unipr.it},
Luca Del Freo, Francesca Terenziani}
\address{Dipartimento di Chimica Generale ed Inorganica, Chimica Analitica 
e Chimica Fisica,\\
Universit\`{a} di Parma, I--43100 Parma, Italy}

\date{\today}

\maketitle

\begin{abstract}
Two-photon absorption (TPA) spectra of push-pull chromophores are
described in terms of a two-state model accounting for 
electron-vibration coupling. 
Vibrations have two main effects in TPA spectra. 
The most obvious one is the appearance of a vibrational structure 
in the spectrum; in this respect we find large Herzberg-Teller contributions.
The second effect was not recognized so far: vibrational
states contribute a new channel to  TPA process, that shows up with
a blue-shift and a distortion of the spectrum. Vibrational-channel
contributions to other NLO responses are shortly discussed.

\end{abstract}

\pacs
 \pagebreak

\section{Introduction}

The large non-linear optical (NLO) responses that characterize 
$\pi$-conjugated materials make them good candidates for innovative 
applications. Among other properties, large two-photon absorption (TPA)
cross-sections are promising for different applications, ranging from 
optical storage to biological optical imaging \cite{albota}. 
As a consequence, both experimental and theoretical work 
\cite{beljonne,kogej,rumi,macak1,macak2,abbotto} 
have received strong impetus  and several
interesting strategies have been proposed to improve NLO responses in
general \cite{marder1,bredas} and TPA cross-section in particular 
\cite{albota,marder2}.
A presently highly debated topic concerns the importance of 
vibrational contributions to NLO responses of organic materials 
\cite{zerbi,bishop1,mukamel,bishop2}.
The coupling of electronic and vibrational degrees of freedom  
(e-ph coupling)  is very important in conjugated 
molecules and polymers \cite{soos} and  is expected to originate large 
 effects in systems with intrinsically non-linear behavior 
\cite{cplanna,cp,freo}. 
General agreement
is presently emerging about the importance of vibronic 
contributions to static NLO responses in conjugated materials 
\cite{zerbi,bishop1,cplanna,cp,freo}. 
This can be easily understood:
the static applied electric fields affect the electronic properties of 
the chromophore by affecting its $\pi$-electron distribution. 
Due to e-ph coupling, the applied fields then also act on the
vibrational states
by displacing nuclei, and this in turn reflects on the electronic
distribution, in a feedback mechanism that  amplifies 
non-linearity \cite{cplanna}. 

The situation is much more complex at 
optical frequencies. A fairly common statement is that nuclei are too 
massive to respond at optical frequencies, so that their contribution to
NLO responses is irrelevant.
In a recent paper \cite{mukamel}, negligible vibrational contributions to 
non-resonant NLO responses have been demonstrated.
Whereas  for actual molecules with 
electronic excitation energies of the order of 1-2 eV and vibrational energies
of about 0.1-0.2 eV it is difficult to find out a frequency range
far enough from both electronic and vibrational resonances as to
safely apply the approximations in Ref.\cite{mukamel},  one indeed does
not expect really large  effects of e-ph coupling in  truly non-resonant NLO 
responses \cite{bishop2}.
As far as resonant processes are concerned, however, the above statement 
does not apply. Quite obviously it leads to erroneous conclusions if applied to
incoherent measurements, i.e. to experiments where a `prepared' state
is allowed to relax for some time before further processing. Just as an
example, we have already proved large and non-trivial effects
of the coupling between 
electronic and slow degrees of freedom in time-resolved
fluorescence \cite{cpl} and femtosecond hole-burning spectra \cite{c102} 
of push-pull chromophores. 
But also in coherent resonant experiments (one- and two-photon absorption
 being simple examples) e-ph coupling is important.  
Just as in one-photon absorption (OPA) spectra, also 
in non-linear spectra  e-ph coupling is 
responsible for  structured, vibronically resolved  band-shapes.
 Large effects of  e-ph coupling have been recognized within the
Condon approximation in TPA spectra of push-pull chromophores \cite{beljonne},
and, within the same approximation, TPA, third harmonic generation (THG) 
and non-degenerate four-wave mixing spectra of the 
conjugated polymer polydiacetylene have been shown to be non-trivially 
affected by e-ph coupling \cite{soos}. 
Recent ab initio calculations
of TPA spectra of a few conjugated molecules \cite{macak1,macak2} 
showed sizable and, in some cases even large,
Herzberg-Teller (HT) corrections, demonstrating once more
the non-trivial role of e-ph coupling in conjugated systems. 

The effects of e-ph coupling in NLO spectra calculated within 
the  Condon and HT approximations are important and have to be 
included in any reliable spectral simulation. However they are not
dissimilar in principle from standard effects in linear absorption spectra.
In fact, within Condon and HT schemes, e-ph coupling only leads to 
a partitioning of the oscillator strength of a given electronic transition 
into several vibronic states, or occasionally (within HT) to a borrowing
of intensity from an allowed transition to a forbidden one. 
Of course the spectral consequences of these well-known phenomena will
be more and more complex and  important the higher is the
order of non-linearity, and the larger is the number of coupled
electronic states involved in the process \cite{soos}. 
But  in non-linear spectra e-ph coupling can 
have much more peculiar effects with no counterpart 
in linear absorption. In conjugated materials e-ph coupling is responsible 
for the appearance in infrared and/or Raman spectra of vibrational
bands with large (sometimes really huge) intensities \cite{zerbi,soos,prb}. 
Then there are {\it vibrational} states, to mean
states within the ground state 
vibrational manifold, that have large transition 
dipole moments and/or transition (Raman) polarizabilities. 
These states, if properly included in sum over states (SOS) expressions 
of NLO susceptibilities, 
provide an important {\it purely vibrational channel} to NLO responses,
that adds to  the standard {\it electronic channel}. 
Vibrational channel contributions to resonant NLO processes have not
been recognized so far: here we demonstrate they can strongly affect
the spectral properties of $\pi$-conjugated systems.

In the next Section we will concentrate on TPA spectra of push-pull
chromophores described in terms of the simplest relevant model, 
the DA dimer with Holstein coupling \cite{cplanna}. 
Within this model we are able to 
calculate numerically exact TPA spectra. We can then discuss the reliability of
Condon and HT approximation schemes. We find sizable HT corrections to
TPA spectra even for systems whose OPA spectra are well reproduced
within the Condon approximation. Even more importantly, we demonstrate large
contributions of  the vibrational channel to TPA spectra.
These  contributions become dominant in the cyanine limit.
In the discussion Section we extend the treatment to other non-linear 
techniques and shortly address the behavior of different 
$\pi$-conjugated systems.

\section{TPA spectra of the DA dimer with Holstein coupling}

The simplest relevant model to describe (low-energy) spectral properties
of push-pull chromophores accounts for just two electronic states,
$|DA\rangle$  and $|D^+ A^-\rangle$, linearly coupled to a vibration,
according to the following Hamiltonian ($\hbar =1$) \cite{cplanna}:

\begin{equation}
  H =  z_0 (1-\sigma _z) -\sqrt{2} t \sigma _x
 -\sqrt{2\epsilon}\omega_v Q \hat \rho + \frac{1}{2}(\omega_v^2Q^2+P^2)
\label{hamiltonian}
\end{equation}
where $Q$ and $P$ are the coordinate and momentum for a harmonic
vibration with frequency $\omega _v$;
$\sigma _z$ and $\sigma _x$ are the Pauli spin operators; $2z_0$ 
is the energy gap at $Q=0$; $\hat \rho = (1-\sigma _z)/2$ is
the ionicity operator measuring the amount of charge separation.
The relaxation energy of $|D^+A^-\rangle$, $\epsilon$,
measures the strength of the e-ph coupling.
The above Hamiltonian assigns the two non-interacting ($\sqrt{2} t =0$) states
two harmonic potential energy surfaces (PES), as shown in Fig. 1a. 
Following Mulliken, we assume orthogonal basis states and neglect all
dipole moment matrix elements but 
$\mu_0 =\langle D^+A^-|\hat \mu| D^+A^-\rangle$, so that 
 the dipole moment operator can be written as:
$\hat \mu = \mu _0 \hat \rho$ \cite{cplanna}.

When the interaction is switched on (without loss of generality we
 set $\sqrt{2} t =1$) the two states mix, and, in the adiabatic approximation,
the ground and excited state PES, shown 
in Fig. 1b, are easily obtained by diagonalizing the relevant 
two-dimensional matrix as a function of $Q$. 
As a consequence of e-ph coupling, i.e. of the relative
displacement of the PES in Fig.~1a, the energy gap 
between the two basis states varies  with $Q$,
so that the amount of mixing between the two states, measured by 
$\rho =\langle  \hat \rho \rangle$, also depends on $Q$. 
Then, both the ground ($|G\rangle$) and excited state ($|E\rangle$)
 wavefunctions depend on $Q$, 
and analytical expressions for the $Q$-dependent properties (energies, 
dipole moments, etc.) are easily obtained. 
In particular the energies of  both states show a non-parabolic
dependence on $Q$, i.e. the two corresponding PES are anharmonic (Fig. 1b). 
This anharmonicity induced
by e-ph coupling has large effects on static NLO responses \cite{cplfreo}, 
where high order $Q$-derivatives of the ground state PES  are involved.
It  also strongly affects incoherent spectral 
measurements where the system tests large regions of the ground
and/or excited state PES \cite{synmet}. 
On the opposite, anharmonicity is not expected
to show up in vertical processes \cite{cplfreo} like OPA or TPA.

In any case, the analytical expressions for the ground and excited state
PES allow for an exact solution of the corresponding vibrational problems.
A detailed description of the adiabatic solution of the
Hamiltonian in Eq.~(1) is deferred to a subsequent publication 
\cite{cinetico}.
Basically, chosen a harmonic phonon basis to describe the vibrational problem
on either surfaces, the corresponding Hamiltonian matrices are
easily written down by expanding the relevant electronic energies 
in powers of $Q$.
Numerically exact vibrational eigenstates are then obtained provided 
that  both the phonon basis and the $Q$-expansion are truncated at a 
sufficiently high order as to get convergence on the evaluated properties.
Similarly, the matrix elements of
$\hat \mu$ on the complete adiabatic basis, are readily obtained
from the $Q$-expansion of the relevant electronic dipole moments 
(either $\mu_G(Q) =\langle G|\hat \mu |G \rangle $, 
or $\mu_E(Q) =\langle E|\hat \mu |E \rangle $ or 
$\mu_{CT}(Q) =\langle G|\hat \mu |E \rangle   $)
closed on the relevant exact anharmonic vibrational eigenstates.
For the calculations presented in this paper,
we chose to describe the vibrational problem in either the ground or
excited state PES on the basis of the eigenstates
of a harmonic oscillator with frequency $\omega_v$, centered at the
minimum of the ground state PES. 
With this choice we found that $\sim 20$ basis 
states and a similar number of terms in the $Q$-expansion of
the electronic energies and/or dipole moments are more than enough
to get convergence. We underline that the  adiabatic approximation 
we adopt in the present treatment is very useful since it  allows for
a clear separation between states in the $G$- and $E$- manifolds, 
and then to distinguish between {\it vibrational} and {\it vibronic}
excitations. The two kinds of states are completely mixed up in a
truly non-adiabatic system.
In any case, as far as the results presented in this paper are concerned,
the  reliability of the adiabatic approximation has been
confirmed by the comparison with exact non-adiabatic results,
obtained along the lines presented in Ref. \cite{cplfreo}.

OPA spectra are calculated as $\omega\,{\rm Im}[ \chi^{(1) }(\omega)]$, with 
 the complex linear susceptibility, $\chi^{(1)}(\omega)$,  defined as:

\begin{equation}
\chi^{(1)}(\omega) = 
      \sum_f \langle g|\hat \mu |f\rangle \langle f|\hat \mu |g\rangle
 \left( \frac{1}{\Omega_{fg}-\omega}+\frac{1}{\Omega^*_{fg}+\omega}\right)
\label{chiopa}
\end{equation}
where $|g\rangle =|G\rangle|0\rangle$ is the global ground state, 
$|f\rangle$ is an excited state, 
$\Omega_{fg}= \omega_{fg} -i\Gamma_f$
is the complex frequency of the $f \leftarrow g$ transition, defined in terms
 of the real transition frequency ($ \omega_{fg}$) and of the damping factor
of the $f$ state ($\Gamma_f$) \cite{orr}. 
The summation in the above equation runs on 
all excitations, including {\it vibrational} excitations (in the 
$G$-manifold)
as well as {\it vibronic} excitations (in the $E$-manifold).
The sum on vibrational excitations describes of course the infrared (IR) 
absorption spectrum, and basically does not contribute to the 
electronic spectrum, dominated by terms that can be
written as $ \langle 0|\mu_{CT}(Q) |v\rangle \langle v| \mu_{CT}(Q) |0\rangle/
(\omega_{CT,v}-\omega -i\Gamma)$, where $|0\rangle$ is the lowest
exact anharmonic vibrational state,  $|v\rangle$ is the $v$-th 
exact anharmonic vibrational state in the $E$-manifold,
and $\omega_{CT,v}$ is the 
$|G\rangle |0\rangle \rightarrow |E\rangle|v\rangle $ transition frequency.

In the following, we fix $\epsilon =1$, and $\omega_v = 0.1$ 
as dimensionless parameters (in $\sqrt{2}t$ units).
In the specific case of  push-pull chromophores, a typical value for
$\sqrt{2}t$ is $\sim$1~eV \cite{cplanna,cpl,jpca2}, 
so that all the energy values can be approximately read in eV.
We present results obtained for three different $z_0$ values,
 to model chromophores
with a very low polarity ($z_0=0.95$, $\rho=0.2$), with intermediate polarity
($z_0=0.66$, $\rho=0.35$) and just in the cyanine limit 
($z_0=0.50$, $\rho=0.5$).
We notice that the properties of the system are symmetric with respect
to the substitution $\rho \rightarrow 1-\rho$.
The upper panels in Fig.~2 report one-photon absorption (OPA) spectra:
exact spectra (continuous lines) are compared to those obtained 
in the Condon (dotted lines) and HT (dashed lines) approximation schemes. 
Both Condon and HT schemes work within the harmonic approximation, 
and  we calculate Condon and HT spectra in the best harmonic approximation 
\cite{cplfreo} for the ground and excited state  PES, i.e. by truncating the 
expansion of  the corresponding potential energies, around the ground state
equilibrium position,
up to the second order in $Q$. In  the Condon approximation, 
the expansion of $\mu_{CT}(Q)$ is truncated at
 the zeroth order (constant) term, whereas in
the HT approximation  the linear term 
in the expansion of the numerator of $\chi^{(1)}$ is also accounted for.
As it turns out from Fig. 2, both Condon and HT approximations are 
fairly good for OPA 
spectra. The minor deviations of HT from Condon spectra
 are  irrelevant in the fairly
broad spectra usually observed for polar chromophores in solution. 
Quite predictably, the harmonic approximation works well to describe 
vertical electronic processes \cite{cplfreo}, and HT (and higher order)
corrections to Condon spectra turn out small for allowed transitions.

Single beam TPA spectra can be calculated as 
$\omega \,{\rm Im}[{\chi_{TPA}(-\omega; \omega,\omega,-\omega)}]$,
where $\chi_{TPA}(-\omega; \omega,\omega,-\omega)$ is the TPA component
of the third order complex susceptibility \cite{orr}:

\begin{eqnarray}
\lefteqn{\chi_{TPA}(-\omega; \omega,\omega,-\omega) \propto 
	\sum_{m,f,n} \langle g|\hat \mu|m\rangle
              \langle m|\overline{\mu}|f\rangle
               \langle f|\overline{\mu}|n\rangle
                \langle n|\hat \mu|g\rangle \times} \nonumber \\
& & \qquad \left[ \frac{1}{(\Omega_{mg}-\omega)(\Omega_{fg}-2\omega)
(\Omega_{ng}-\omega)} 
+\frac{1}{(\Omega^*_{mg}+\omega)(\Omega_{fg}-2\omega)
(\Omega_{ng}-\omega)}\right. \nonumber \\
& & \qquad \left. +\frac{1}{(\Omega^*_{mg}+\omega)(\Omega^*_{fg}+2\omega)
(\Omega_{ng}-\omega)}
+\frac{1}{(\Omega^*_{mg}+\omega)(\Omega^*_{fg}+2\omega)
(\Omega^*_{ng}+\omega)}\right]
\label{chitpa}
\end{eqnarray}
where the triple sum runs on all excited states. 
Of course, if we are interested
in TPA processes leading to the $E$-manifold, 
(i.e.  processes measured for $\omega \sim \omega_{CT}/2$),
then $f$ only counts vibronic excitations.
But $m$ and $n$ run on both {\it vibronic} and {\it vibrational}
excitations.

In all available studies of vibrational contributions to TPA spectra 
\cite{beljonne,macak1,macak2,soos},
all summations in Eq.~(\ref{chitpa}) have been limited to vibronic
excitations only, i.e. only the  electronic channel
to TPA has been accounted for. 
In this approximation, $\chi_{TPA}$ reduces to:
\begin{eqnarray}
\lefteqn{\chi_{TPA}^{el}(-\omega; \omega,\omega,-\omega) \propto} \nonumber \\
& &\qquad \sum_{u,v,w}   \langle 0|\mu_{CT}(Q)|u\rangle
              \langle u |\mu_E(Q)-\mu|v\rangle
               \langle v|\mu_E(Q)-\mu|w\rangle
                \langle w| \mu_{CT}(Q)|0\rangle \times\nonumber \\
& & \qquad \left[\frac{1}{(\Omega_{CT,u}-\omega)(\Omega_{CT,v}-2\omega)
(\Omega_{CT,w}-\omega)}
+\frac{1}{(\Omega_{CT,u}^*+\omega)(\Omega_{CT,v}-2\omega)
(\Omega_{CT,w}-\omega)} \right.
\nonumber \\ 
& & \qquad \left. +\frac{1}{(\Omega_{CT,u}^*+\omega)(\Omega_{CT,v}^*+2\omega)
(\Omega_{CT,w}-\omega)}
+\frac{1}{(\Omega_{CT,u}^*+\omega)(\Omega_{CT,v}^*+2\omega)
(\Omega_{CT,w}^*+\omega)}\right] 
\label{chitpael}   
\end{eqnarray}
where $u,v$ and $w$ 
run on vibrational states in the $E$-manifold,
and $\mu =\langle 0|\mu_G(Q)|0\rangle $ 
is the ($Q$-independent) ground state dipole moment. 
In the middle panels of Fig. 2 we report 
the electronic channel contribution to TPA spectra, calculated
for the exact adiabatic eigenstates of the Hamiltonian in Eq.~(1)
(continuous lines). In the same figure, dotted and dashed
lines refer to spectra obtained within the Condon and HT approximation schemes.

Even accounting for just the electronic channel to TPA absorption,
we can already stress the importance of a proper modeling of
 e-ph coupling in
TPA spectra. If e-ph coupling is not accounted for, 
both OPA and TPA spectra have a single Lorentzian shape, and
are exactly superimposed if the TPA frequency scale is expanded by 
a factor of 2 with respect to the OPA scale. At the Condon level,
the situation is not very different: 
e-ph coupling originates structured bands in both OPA and TPA spectra,
but with  similar shape, so that the two spectra are 
essentially superimposable as in the absence of coupling. 
An obvious deviation is observed at  $\rho=0.5$, where, in the harmonic
approximation, the TPA spectrum exactly vanishes due to 
the factor $\mu_{E}-\mu$, proportional to $1-2\rho$ 
in the dimer model.
At variance with OPA spectra, however,
HT corrections are very important in TPA,
 at least for not too neutral molecules.
For $\rho =0.35$, the HT spectrum in the central panel of Fig.2 
has a qualitatively different shape 
from the Condon spectrum, with the 0-1 line acquiring intensity at the expense
of the 0-0 line. 
Apart from the pathological $\rho=0.5$ case, the HT approach
offers a good approximation to the electronic channel contribution to TPA,
confirming the reliability of the harmonic approximation 
to describe vertical electronic processes \cite{cplfreo}.

Up to now we have only considered the electronic channel contribution to TPA
process, where  e-ph coupling mainly contributes in defining vibronically 
structured band-profiles. We already got a fairly impressive result:
the Condon approximation to TPA spectra is in general inadequate, even
in cases where it works well for OPA spectra.
However in TPA spectra, and more generally in non-linear spectra, 
vibrations  have a much more important role than simply structuring the 
observed band-shapes. They in fact open a vibrational channel
to non-linear processes. The lower panels in Fig. 2 show the total TPA 
spectrum, obtained from the imaginary part of $\chi_{TPA}$
 in Eq.~(\ref{chitpa}), by allowing $m$ and $n$ indices to run on both
{\it vibrational} and {\it vibronic}
 excitations. The comparison between continuous lines in the middle and lowest 
panels in Fig.2 clearly demonstrates the importance of the vibrational 
channel to TPA process, with sizable effects on the absorption frequency, on 
the band-shape as well as on the total intensity. 
Overall, the vibrational channel 
increases the intensity of higher vibronic replicas at the expense
of the lowest replicas, with effects that are most important at $\rho \sim 0.4$.
This implies a blue-shift of the TPA spectrum if compared with the rescaled
OPA spectrum.  In the extreme case of $\rho=0.5$, the blue-shift amounts
to a vibrational spacing. Intensity effects are negligible for $\rho \sim 0.2$,
but at $\rho=0.35$ the vibrational channel amplifies TPA response by a 
1.2 factor, that becomes a  factor of 4 in the  $\rho=0.5$ case.

\section{Discussion}

Vibrational channel contributions to TPA process arise from terms in 
Eq.~(\ref{chitpa}) where $m$ or $n$, or both $m$ and $n$ run on vibrational 
excitations, so that in each term contributing to the vibrational channel 
at least a factor like $\langle 0 |\mu_G(Q)|v\rangle$ appears,
where  $v$ is one of the {\it vibrational} states in the $G$-manifold. 
The square of this term measures
the absorption intensity of the $v$-th vibrational state.  
Since the fundamental vibration has 
by far the largest IR intensity \cite{cplfreo}, 
the vibrational channel to TPA is 
dominated by it. A first approximation to $\chi_{TPA}^{vib}$ is then:

\begin{eqnarray}
 \chi_{TPA}^{vib} 
&\sim & \left[
\sum_{v,w}2
\frac{\mu_{IR}\langle 1|\mu_{CT}(Q)|v\rangle \langle v|\mu_E(Q)-\mu|w\rangle
\langle w| \mu_{CT}(Q)|0\rangle}
{(\omega_v-\omega)(\Omega_{CT,v}-2\omega)(\Omega_{CT,w}-\omega)}
\right. \nonumber \\
& &  \left.
+\sum_{v}
\frac{\mu_{IR}\langle 1|\mu_{CT}(Q)|v\rangle \langle v|\mu_{CT}(Q)|1\rangle
\mu_{IR}}
{(\omega_v-\omega)(\Omega_{CT,v}-2\omega)(\omega_v-\omega)}\right]
\label{chitpavib}
\end{eqnarray}
where $\mu_{IR}=\langle 0 |\mu_G(Q)|1\rangle$ is the dipole moment of 
the fundamental vibrational transition.
The relative importance of $\chi_{TPA}^{vib}$ vs $\chi_{TPA}^{el}$
goes as $(\mu_{IR}/(\mu_E-\mu)) (|\omega_{CT}-\omega|/|\omega_v -\omega|)$
or as its square, depending if the first or the second term in the
above equation is concerned.
Since $\mu_{IR} \propto [\rho(1-\rho)]^{3/2}$ \cite{jpca2} maximizes at 
$\rho=0.5$, whereas  $\mu_E-\mu  \propto (1-2\rho)$ vanishes there, 
the relative importance of $\chi_{TPA}^{vib}$ increases fast towards 
the cyanine limit. 
On the other hand, the frequency of vibrational modes, 
$\omega_v \sim 0.1-0.2$ eV, is not far from the laser frequency 
$\omega \sim \omega_{CT}/2 \sim 0.5-1.5$ eV,
so that the ratio $((\omega_{CT}-\omega)/(\omega_v -\omega))$ enhances 
the vibrational contribution over  the electronic
one by a factor ranging up to 3.

The solution of the two-state model in the adiabatic approximation
allows for reliable estimates of the relative 
magnitude of electronic and vibrational contributions to TPA spectra 
and even more importantly offers a natural understanding of 
the physical origin of the different contributions. 
An obvious drawback of the two-state model
is however the impossibility to account for the role of highly excited states.
Of course, extending the sums in Eq.~(\ref{chitpa}) to excited states ($E'$)
higher than the lowest one ($E$) opens additional channels to the TPA process.
These $E'$-channel contributions to TPA have $m$ or $n$ or both $m$ and $n$
in Eq.~(3) running over vibronic states pertaining to the $E'$-manifolds.
By closing over the vibrational states of these  high-energy levels, 
one can roughly  estimate the magnitude of 
$E'$-channel contributions in terms of
the $G\rightarrow E'$ and $E\rightarrow E'$ transition dipole moments 
(beside of $G\rightarrow E$) and of the corresponding transition energies.
At the best of our knowledge, experimental estimates of the $E \rightarrow E'$
transition dipole moments are only available for coumarin 153 dispersed
in CH$_3$CN \cite{kovalenko}. For this molecule, excited state
transient absorption measurements  yield a $E \rightarrow E'$ transition
dipole moment of $\sim 10$~D. Transition dipole moments of similar 
magnitude are also evaluated, from OPA spectra, for the two transitions
 $G\rightarrow E$ and $G\rightarrow E'$, occurring at energies
of $\sim 3.0$~eV and $\sim 5.6$~eV, respectively \cite{kovalenko,c153}.
The joined analysis of electronic  and vibrational spectra of
the same molecule \cite{c153} yields
 $\rho \sim 0.23$, $\epsilon \sim 1$~eV
and $\omega_v \sim 0.2$~eV, in fairly good agreement with previous estimates
on a similar dye (coumarin 102) \cite{cpl}. 
Based on these data, the vibrational and $E'$-channel contributions 
to TPA spectra turn out of the same order of magnitude. 
In view of the fairly low polarity of coumarin 153 and of its fairly
high $\omega_{CT}$ frequency, we believe that for most push-pull chromophores
the vibrational channel contribution to TPA spectra is larger than
$E'$-channel contributions.
In fact, preliminary ZINDO calculations on a few chromophores  
confirm  that, at least for chromophores with intermediate polarity
($\rho \sim 0.3-0.7$), the vibrational contributions to TPA spectra
are important and cannot be disregarded with respect to $E'$-channel 
contributions. 
Refined models for NLO responses accounting for plenty of excited states
are bound to fail if no proper account is given for the 
contributions of vibrational states.

Besides TPA spectra, other NLO responses 
can be affected by the vibrational channel. Fig. 3 shows 
the amplitude of the third harmonic generation (THG) signal in the region
$\omega\sim\omega_{CT}/3$, for three different $\rho$ values. 
Continuous lines refer to the total response, obtained by allowing 
the triple sums in $\chi_{THG} =\chi^{(3)}(-3\omega;\omega,\omega,\omega)$ 
to run on both vibronic and vibrational excitations.
Dashed lines instead  account for the electronic channel 
to THG, obtained by summing only on  vibronic states.
The vibrational channel strongly affects  THG response, with largest
effects at $\rho \sim 0.3$, where the response is amplified by a factor
$\sim 2$ in the shown spectral region. 
Calculations are performed using the complete expression for
$\chi_{THG}$, but, in the portion of the spectrum shown in the 
figure, the terms  with a resonating denominator are clearly dominant, 
so that we can approximate $\chi_{THG}$ as follows:

\begin{eqnarray}
\lefteqn{ \chi_{THG}(-3\omega; \omega,\omega,\omega) \sim}  \nonumber \\
& &	\left[ \sum_{f,m,n} \frac
            {\langle g|\hat \mu|f\rangle
              \langle f|\overline { \mu} |m\rangle
               \langle m|\overline { \mu }|n\rangle
                \langle n|\hat \mu|g\rangle }
           {(\Omega_{fg}-3\omega)(\Omega_{mg}-2\omega)(\Omega_{ng}-\omega)}
      -\sum_{f,m} \frac{\langle g|\hat \mu|f\rangle
              \langle f|\hat \mu|g\rangle
               \langle g|\hat \mu|m\rangle
                \langle m|\hat \mu|g\rangle }
           {(\Omega_{fg}-3\omega)(\Omega_{fg}-\omega)(\Omega_{mg}-\omega)}
\right]
\label{chithg}
\end{eqnarray}
where $f$ runs on vibronic excitations, just to guarantee an
almost  resonant denominator, $\omega_{CT}-3\omega$,
and $m, n$ run on both vibrational and vibronic excitations. 
 The vibrational channel contribution to THG is more complex than
for TPA. A few terms of $\chi_{THG}^{vib}$, obtained by allowing $n$ or $m$
or both $m$ and $n$ in the above equation to run on vibrational excitations, 
contain at the numerator $\mu_{IR}$ or its square. A term however has a
different  numerator: $\sum_{v,w}
\langle 0| \mu_{CT}(Q)|v\rangle \langle v|\mu_{CT}(Q)|1\rangle
\langle 1| \mu_{CT}(Q)|w\rangle \langle w|\mu_{CT}(Q)|0\rangle$,
that is related to the non-resonant Raman intensity of the fundamental
vibration. It is difficult to disentangle the 
contributions of the various terms to 
the observed signal, but the importance of Raman modes
is suggested by the observation of the maximum 
vibrational contribution  to THG just at $\rho \sim 0.3$, 
where the non-resonant Raman intensity maximizes for the DA dimer model
\cite{jpca1}.

Having rationalized the vibrational channel contributions to TPA and THG 
spectra of the DA dimer model in terms of IR and non-resonant Raman intensity 
of coupled vibrations, we can shortly extend our discussion to different
systems. In conjugated polymers the coupled vibrations are not IR active,
whereas they have huge (non-resonant) Raman intensities \cite{zerbi,soos}. 
Then we expect no contribution from the vibrational channel in TPA
spectra of pristine polymers, whereas the vibrational contributions to 
THG spectra can be important. 
A detailed description of NLO properties of polymers is difficult,
in view of their fairly crowded excitation spectra. 
Preliminary results on a simple relevant model, the
self-dimer with Peierls coupling \cite{smanna},
confirm large vibrational channel contributions to THG
spectra, and negligible contributions to TPA spectra.
Of course large vibrational contributions to TPA spectra can be
expected for doped or photoexcited polymer samples, based on the
very large IR intensities associated with charged solitonic or polaronic
defects \cite{soos,prb}.

\section{Conclusions}

In conjugated molecules and polymers e-ph coupling is very effective.
It shows up with largely anharmonic PES, that, in conjugated polymers,
often show 
more than a single minimum \cite{mazumdar} to signal the Peierls instability. 
A second fairly ubiquitous 
consequence of e-ph coupling is the large nuclear displacement 
upon electronic excitation, with well apparent effects in absorption 
and resonant Raman spectra \cite{bredas,soos,cpl,jpca2,jpca1}. 
In systems with large e-ph coupling,
static electric fields can draw nuclei 
out of equilibrium, and  this is the basic reason why e-ph
coupling is so effective in amplifying static  NLO responses via anharmonic
contributions \cite{cplfreo}.
Fields at optical frequencies are too fast to displace  nuclei,
but in incoherent measurements nuclei are allowed to relax
for a while following the application of the preparation pulses: 
once more e-ph coupling is important in defining the shape of the 
involved PES \cite{c102,synmet}.
Electron-phonon coupling largely affects vibrational spectra, 
with a softening of the
relevant modes and a large increase of infrared and/or Raman intensities.
Therefore in  SOS expressions for the susceptibilities, 
vibrational states can give  large contributions.
For static susceptibilities this contribution 
indeed accounts for the anharmonicity of the ground state PES 
\cite{cplfreo,cinetico}.

As far as resonant processes are concerned, the importance of the vibrational
contributions has not been
 fully appreciated so far. Here we have analyzed TPA spectra of the DA
dimer model --- the simplest model relevant to push-pull chromophores ---
and proved that the vibrational contribution is very important in defining
the shape, frequency and intensity of the signal.
Including vibrational degrees of freedom has two major effects in TPA spectra,
and more generally in resonant NLO processes.
The first, fairly obvious, effect is the appearance of a vibronic structure
in TPA spectra. This effect was already recognized 
\cite{beljonne,macak2,bredas,soos}; here
we prove the validity of the harmonic approximation in this context, but the 
inadequacy of the Condon approximation. Specifically, HT
corrections to TPA spectra are important even in cases when OPA spectra are
 well reproduced within the Condon approximation.
The second effect of vibrations is more interesting and has not been recognized
 so far. Vibrational states in the ground state manifold in fact contribute 
a new and important channel to TPA process. 
The vibrational channel contribution shows up with a blue-shift
of the TPA spectrum with respect to the rescaled OPA spectrum and with a
large distortion of the TPA band-shape.
These  contributions are related to simple 
spectroscopic observables as inferred from vibrational and linear 
absorption spectra.
This immediately leads to safe predictions about other systems. As a matter
of fact, vibrational channel contributions to TPA spectra are related to
the IR intensity of coupled modes, so that they are expected to be negligible
in any polymer (but possibly dominant in doped and/or photoexcited polymers).
THG spectra instead have contributions from Raman modes, so that
large effects of e-ph coupling  are expected  in THG spectra of conjugated
polymers.

The behavior of conjugated materials can only
be understood by properly accounting for e-ph coupling, that affects in 
different ways and for different reasons the spectral properties of
these systems. Common beliefs based on the behavior of other systems
where e-ph coupling is not so effective, and/or extrapolated from our
knowledge of their 
linear spectroscopic behavior need to be carefully reconsidered.
A joint experimental and theoretical effort is needed to extract the
correct physics out of these fascinating materials.

\vspace{0.7cm}

\centerline { \bf {ACKNOWLEDGMENTS}}
\vspace{0.5cm}
We thank A.~Girlando for  useful discussions.
Work supported by the Italian  National Research Council (CNR)
within its ``Progetto Finalizzato Materiali Speciali per Tecnologie
Avanzate II'',   
and by the Ministry of University and of Scientific
and Technological Research (MURST).

 \vfill
 \eject

 \vfill
 \eject

\pagebreak

\begin{figure}
\begin{center}
\mbox{\epsfig{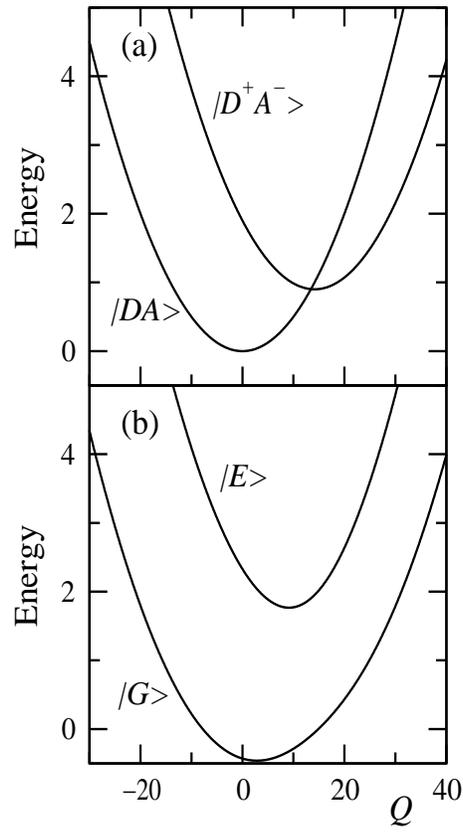}}
\end{center}
\caption{
Potential energy surfaces sketched along the vibrational
 coordinate $Q$
for (a) the basis states ($\sqrt{2}t=0$), and (b) the exact eigenstates
($\sqrt{2}t=1$). Both panels refer to $z_0=0.95$ and $\epsilon =1$.}
\end{figure}

\begin{figure}
\begin{center}
\mbox{\epsfig{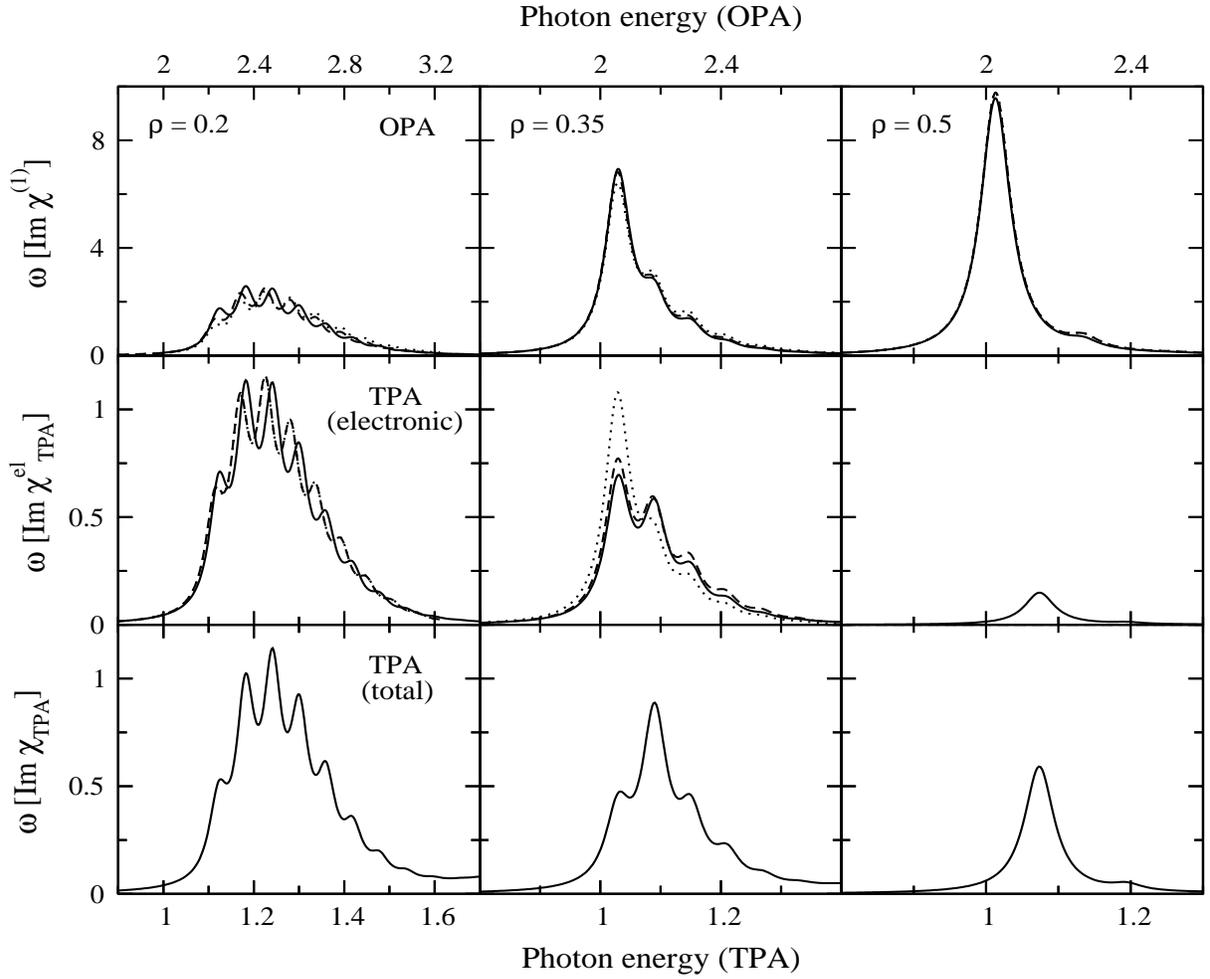}}
\end{center}
\caption{
OPA and TPA spectra calculated for $\epsilon=1$ and $\omega_v = 0.1$
in units of $\sqrt{2}t$. Left, middle and right
panels refer to $\rho=0.2$, 0.35, 0.5, respectively.
The vibronic and vibrational damping factors are fixed to 
0.05 and 0.005, respectively.
Upper panels report exact OPA spectra as continuous lines. Dashed and dotted
lines correspond to spectra obtained in the HT and Condon approximation,
respectively. Middle panels report the electronic channel contribution to
TPA spectra; continuous, dashed and dotted lines refer to exact spectra and
to spectra obtained in the HT and Condon approximation, respectively.
Bottom panels show the total TPA spectrum including the vibrational channel
contribution.
The $x$-axis scale at the top and bottom of the figure report the
energy of the OPA and TPA spectra, respectively.
$y$-axes (arbitrary units) preserve the relative intensities.}
\end{figure}

\begin{figure}
\begin{center}
\mbox{\epsfig{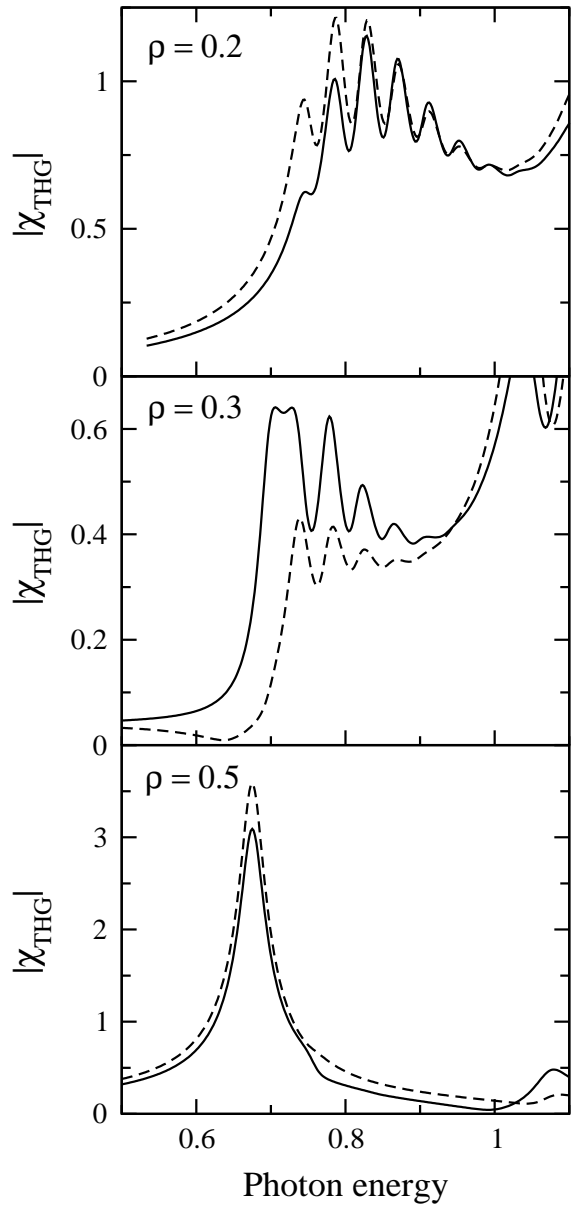}}
\end{center}
\caption{
THG spectra calculated for the same parameters as in Fig.~1.
Upper, middle and bottom
panels refer to $\rho=0.2$, 0.3, 0.5, respectively.
THG spectra including the vibrational
channel contribution (continuous lines) are compared with the purely
electronic THG spectra (dashed lines).}
\end{figure}


\begin{references}
 
\bibitem{albota} 
	M. Albota, D. Beljonne, J.-L. Br\'edas, J.E. Ehrlich, J.-Y. Fu,
	A.A. Heikal, S.E. Hess, T. Kogej, M.D. Levin, S.R. Marder,
	D. McCord-Maughon, J.W. Perry, H. R\"ockel, M. Rumi, G. Subramaniam,
	W.W. Webb, X.-L. Wu, C. Xu, Science 281 (1998) 1653.

\bibitem{beljonne} D. Beljonne, J.-L. Br\'edas, M. Cha, W.E. Torruellas, 
	G.I. Stegeman, J.W. Hofstraat, W.H.G. Horsthuis, G.R. M\"ohlmann,
	J. Chem. Phys. 103 (1995) 7834.

\bibitem{kogej} T. Kogej, D. Beljonne, F. Meyers, J.W. Perry, S.R. Marder,
	J.-L. Br\'edas, Chem. Phys. Lett. 298 (1998) 1.

\bibitem{rumi} M. Rumi, J.E. Ehrlich, A.A. Heikal, J.W. Perry, S. Barlow,
	Z. Hu, D. McCord-Maughon, T.C. Parker, H. R\"ockel, S. Thayumanavan,
	S.R. Marder, D. Beljonne, J.-L. Br\'edas, 
	J. Am. Chem. Soc. 122 (2000) 9500.

\bibitem{macak1} P. Macak, Y. Luo, P. Norman, H. \AA gren,
	J. Chem. Phys. 113 (2000) 7055.

\bibitem{macak2} P. Macak, Y. Luo, H. \AA gren, 
	Chem. Phys. Lett. 330 (2000) 447.

\bibitem{abbotto} A. Abbotto, L. Beverina, R. Bozio, S. Bradamante, 
	C. Ferrante, G. A. Pagani, R. Signorini, Adv. Mater. 12 (2000) 1963.

\bibitem{marder1} S.R. Marder, B. Kippelen, A. K.-Y. Jen, N. Peyghambarian,
	Nature 388 (1997) 845.

\bibitem{bredas} J.-L. Br\'edas, K. Cornil, F. Meyers, D. Beljonne,
	in: T.A. Skotheim, R.L. Elsenbaumer, J.R. Reynolds (Eds.),
	Handbook of Conducting Polymers, 
	Marcel Dekker, New York, 1997, p. 1.

\bibitem{marder2} S.R. Marder, W.E. Torruellas, M. Blanchard-Desce, V. Ricci,
	G.I. Stegeman, S. Gilmour, J.-L. Br\'edas, J. Li, G.U. Bublitz,
	S.G. Boxer, Science 276 (1997) 1233.

\bibitem{zerbi} M. Del Zoppo, C. Castiglioni, P. Zuliani, G. Zerbi,
	in: T.A. Skotheim, R.L. Elsenbaumer, J.R. Reynolds (Eds.),
	Handbook of Conducting Polymers, 	
	Marcel Dekker, New York, 1997, p. 765.

\bibitem{bishop1} D.M. Bishop, Adv. Chem. Phys. 104 (1998) 1.

\bibitem{mukamel} V. Chernyak, S. Tretiak, S. Mukamel, 
	Chem. Phys. Lett. 319 (2000) 261.

\bibitem{bishop2} D.M. Bishop, B. Champagne, B. Kirtman,
	Chem. Phys. Lett. 329 (2000) 329.

\bibitem{soos} Z.G. Soos, D. Mukhopadhyay, A. Girlando, A. Painelli,
	in: T.A. Skotheim, R.L. Elsenbaumer, J.R. Reynolds (Eds.),
	Handbook of Conducting Polymers, 
	Marcel Dekker, New York, 1997, p. 165.

\bibitem{cplanna} A. Painelli, Chem. Phys. Lett. 285 (1998) 352.

\bibitem{cp} A. Painelli, Chem. Phys. 245 (1999) 185;
	A. Painelli, Chem. Phys. 253 (2000) 393.

\bibitem{freo} L. Del Freo, A. Painelli, A. Girlando, Z. G. Soos,
	Synth. Met. 116 (2001) 257.

\bibitem{cpl} A. Painelli, F. Terenziani, Chem. Phys. Lett. 312 (1999) 211.

\bibitem{c102} A. Painelli, F. Terenziani, Synth. Met., in press.

\bibitem{prb} A. Painelli, L. Del Freo, A. Girlando, Z.G. Soos,
	Phys. Rev. B 60 (1999) 8129.

\bibitem{cplfreo} L. Del Freo, A. Painelli, Chem. Phys. Lett., submitted.

\bibitem{synmet} A. Painelli, L. Del Freo, F. Terenziani, 
	Synth. Met, in press.

\bibitem{cinetico}  L. Del Freo, F. Terenziani, A. Painelli, in preparation.

\bibitem{orr} B.J. Orr, J.F. Ward, Molec. Phys. 20 (1971) 513.

\bibitem{jpca2} F. Terenziani, A. Painelli, D. Comoretto,
	J. Phys. Chem. A 104 (2000) 11049.

\bibitem{kovalenko} S.A. Kovalenko, J. Ruthmann, N.P. Ernsting,
	Chem. Phys. Lett. 271 (1997) 40.

\bibitem{c153} F. Terenziani, A. Painelli, unpublished results.

\bibitem{jpca1} A. Painelli, F. Terenziani, J. Phys. Chem. A 104 (2000) 11041.

\bibitem{smanna} A. Painelli, Synth. Met. 101 (1999) 218.

\bibitem{mazumdar} D. Baeriswyl, D.K. Campbell, S. Mazumdar,
	in: H. Kiess (Ed.), Conducting Polymers, 
	Springer-Verlag, Heidelberg, 1992, p. 7.


\end{references}
\end{document}